\documentclass[12pt]{article}
\usepackage{epsfig}

\begin{document}

\title{
{\Large \bf Non-linear    
generalization of the $sl(2)$ algebra
\footnote{Physics Letters A 300 (2002) 205} }
      }

\date{}

\author{
E. M. F. Curado$^1$ and M. A. Rego-Monteiro$^{2}$\\
Centro Brasileiro de Pesquisas F\'\i sicas, \\
Xavier Sigaud 150, 22290-180 - Rio de Janeiro, RJ, Brazil \\
$1$) evaldo@cbpf.br \\
$2$) regomont@cbpf.br}

\maketitle

\begin{abstract}
 
\indent
We present a generalization of the $sl(2)$ algebra where the
algebraic relations are constructed with the help of a general
function of one of the generators. When this function is linear
this algebra is a deformed $sl(2)$ algebra. In the non-linear case,
the finite dimensional representations are constructed in two
different ways. In the first case, which provides finite dimensional
representations only for the non-linear case, these representations
come from solutions to a dynamical equation and we show how to
construct explicitly these representations for a general quadratic
non-linear function. The other type of finite dimensional
representation comes from solutions to a cut condition equation. We
give examples of solutions of this type in the non-linear case as
well.

\end{abstract}

\vspace{1cm}

\begin{tabbing}

\=xxxxxxxxxxxxxxxxxx\= \kill

$su_{q}(2)$ algebra.
 \\


\end{tabbing}

\newpage 

\section{Introduction}

\indent

Two years ago, a class of generalized Heisenberg algebras was
introduced where, within this class, one finds deformed and
non-deformed Heisenberg-type algebras \cite{algebra1,algebra2}.
The representations of these algebras were constructed using concepts
of dynamical systems as attractors and their stabilities
\cite{algebra2}. 
A simple example of this class of algebras is the well-known
$q$-oscillators \cite{macfarlaine}. It was also shown that
this class describes the Heisenberg-type algebraic structure
of a family of one-dimensional quantum systems having any two
successive energy levels related by $\epsilon_{n+1}=
f(\epsilon_n)$, where $f$ is a characteristic function of the
algebra \cite{comhugo}. As a possible physical consequence
of this family of Heisenberg-type algebras, it was shown that it
is possible to construct a non-standard quantum field theory
based on this algebra \cite{tcampos}.

In \cite{physa} it was shown that this way of generalizing an algebraic
structure can be implemented to the $sl(2)$ algebra as well. In
this case the algebraic relations of the generalized $sl(2)$ algebra
also present a characteristic function of one of the generators of the
algebra so that when this function is taken as $f(x) = x - 1$ the
well-known $sl(2)$ algebra is recovered.

In this letter we discuss the finite dimensional representations of
this generalized $sl(2)$ class of algebras. These representations
are constructed by solving an equation that admits two types of
solutions. The first type, which happens only in the non-linear
generalization of $sl(2)$, is obtained by solving a dynamical
equation and the second type of finite dimensional representation
is obtained from the solutions to a cut condition equation. This
last type of solution is responsible for all finite dimensional
representations of the linear case. We also construct a map
connecting the linear case of this generalized $sl(2)$ algebra
with $sl_{q}(2)$. Since the generalized $sl(2)$ class of algebras
presented here recovers a deformed $sl(2)$ algebra for the linear
case which is known to be connected with some simple two-dimensional
integrable systems, \cite{algdeformed} we hope that the non-linear
case of this non-linear generalized algebra can be of help in
understanding more complex integrable systems.

In section II we introduce the iterative generalized $sl(2)$
algebra and present the general conditions to obtain finite and infinite
dimensional representations. In section III, we establish the connection
of 
the linear case of this algebra with the $sl_{q}(2)$ algebraic structure.
We also show that it is possible to study the finite dimensional
representations taking into account the stability of the fixed points of
$f$.
In section IV we study the generalized algebra for a non-linear
function $f$ and show that part of the finite dimensional
representations is obtained by finding the cycles of the characteristic
function of the algebra, $f(x)$. We also construct another type of
finite dimensional representations in this case, which is associated
with the solutions to a cut condition equation.

\section{Iterative generalized $sl(2)$ algebra}

\indent 

Let us consider the following algebraic relations among
the generators $J_{0}$, $J_{+}$ and $J_{-}$:
\begin{eqnarray}
J_{0} \, J_{-} &=& J_{-} \, f(J_{0}) \,\, ,
\label{eq:alg1}\\
J_{+} \, J_{0} &=& f(J_{0}) \, J_{+} \,\, ,
\label{eq:alg2}\\
\left[ J_{+},J_{-} \right] &=& J_{0} (J_0 + 1) -
f(J_{0}) (f(J_{0}) + 1) \;\;\; ,
\label{eq:alg3}
\end{eqnarray}
where we assume $J_- = J_+^{\dagger}$, $J_0^{\dagger} = J_0$ and
that $f(J_0)$ is an analytical function in $J_0$ \footnote{We stress that
the results of this section are valid for a general analytical function
$f$.}. 

This algebra satisfies, for all functions $f$, the Jacobi identity
\begin{equation}
    [J_{0}, [J_{+}, J_{-}]] +  [J_{-}, [J_{0}, J_{+}]] +
     [J_{+}, [J_{-}, J_{0}]] = 0 \, .
    \label{eq:jacobi}
\end{equation}
The first term of the L.H.S. is identically null due to eq. (\ref{eq:alg3}). 
To show that the sum of the other two terms is equal to zero
it is enough to expand them and use the property, derived from
eqs. (\ref{eq:alg1}) and (\ref{eq:alg2}) , that
$[J_{0}, J_{+} \, J_{-}] = 0 $ .

Using the algebraic relations in eqs. (\ref{eq:alg1}-\ref{eq:alg3}) it
can be shown that the operator
\begin{equation}
C = 1/2 \, \left\{J_{+} \, J_{-} + J_{-} \, J_{+} + J_{0}(J_{0} + 1) +
f(J_{0}) (f(J_{0}) + 1) \right\} \;\;\;
\label{eq:casimir} 
\end{equation}
satisfies the commutation relations $\left[ C,J_{0} \right] =
\left[ C,J_{\pm} \right] = 0$ ,
i.e., $C$ is a Casimir operator of the algebra.

If we substitute the specific function $f(J_{0}) = J_{0} - 1$
in eqs. (\ref{eq:alg1}-\ref{eq:alg3}),
we reobtain the well-known $sl(2)$  algebra. Thus, the
algebraic relations proposed in eqs. (\ref{eq:alg1}-\ref{eq:alg3})
contain, as a particular case, the $sl(2)$ algebra when we choose a
specific linear function of $J_{0}$. To discuss this algebra and
the role of the function $f$ in it, we present
its respective representation theory.

Under the hypothesis that the function $f$
and the initial value $\alpha_{j}$ satisfy
$\alpha_{j}>f(\alpha_{j}) > f(f(\alpha_{j})) > \ldots >
f^m(\alpha_{j}) > \ldots$, where $f^m$ means the $m$-th iterate of 
$\alpha_{j}$ through $f$ and $m$ is a positive integer, and that 
there is a vector, the highest weight vector, such that
\begin{equation}
    J_{+} |\alpha_{j}, j  \rangle = 0 \, ,
    \label{eq:j+0}
\end{equation}
we obtain for a general $m$ lying between $0$ and $2j$ \cite{physa}:
\begin{eqnarray}
    J_{0} |\alpha_{j},j-m  \rangle  & = & \alpha_{j-m}
    |\alpha_{j},j-m \rangle
    \label{eq:j0m}  \\
    J_{+} |\alpha_{j},j-m \rangle & = &
    N_{m-1} |\alpha_{j},j-m+1 \rangle
    \label{eq:jm+}  \\
    J_{-} |\alpha_{j},j-m \rangle & = &
    N_{m} |\alpha_{j},j-m-1 \rangle
    \label{eq:jm-}  \\
    C |\alpha_{j}, j-m \rangle & = &
    \alpha_{j} (\alpha_{j}  + 1) |\alpha_{j},j-m \rangle \, ,
    \label{eq:Tm}
\end{eqnarray}
where $N_{m}^2 = (\alpha_{j}  - \alpha_{j-m-1} ) (\alpha_{j}  +
\alpha_{j-m-1}  + 1) = \alpha_{j} (\alpha_{j} +1) -
\alpha_{j-m-1} (\alpha_{j-m-1} +1)$
that can be proved in a similar way to the proof made
for the generalized Heisenberg algebra in \cite{algebra2},
where the additional equation, eq. (\ref{eq:Tm}), is obtained
using eqs. (\ref{eq:alg3}) and (\ref{eq:casimir}).
As shown in \cite{physa} the allowed values of $\alpha_{i}$ 
satisfy
\begin{equation}
    \alpha_{j}  \ge \alpha_{i} \ge \alpha_{b} \,
    (=f^{d-1}(\alpha_{j})) \, .
    \label{eq:rangeb}
\end{equation}

If we put
$m = d-1$ in eq. (\ref{eq:jm-}), in order to have a $d$-dimensional
representation, we must solve
\begin{equation}
N_{d-1}= 0 \, .
\label{eq:geral}
\end{equation}
This equation is satisfied if $\alpha_{j} = \alpha_{j-d}$ or if
the highest weight of the representation satisfies a cut condition
equation
\begin{equation}
\alpha_{j}  + \alpha_{j-d} + 1 = 0 \, .
\label{eq:corte2}
\end{equation} 
As it is clear in what follows, the equation $\alpha_{j} =
\alpha_{j-d}$
will have solutions only in the non-linear case (for $d>1$) while the finite
dimensional representations coming from solutions to eq.
(\ref{eq:corte2})
will have highest weights satisfying
 \begin{equation}
  \alpha_{j} > - \alpha_{j}  - 1 \Rightarrow \alpha_{j} > -1/2 \, ,
     \label{eq:alfameio}
 \end{equation}
since $\alpha_{j} > \alpha_{j-d}$.

%

\section{Linear functions}
\label{sec.linear}

\subsection{$sl(2)$}\label{subsec.su2}

\indent

As we have seen, with $f(J_{0}) = J_{0} - 1$ we reproduce the
commutation 
relations of the $sl(2)$ algebra. It is straightforward to verify that
for the function under consideration a general eigenvalue $\alpha_{j-m}$
can be written as:
\begin{equation}
    \alpha_{j-m} = f^{m}(\alpha_{j}) = \alpha_{j} -m \, .
    \label{eq:fsu2}
\end{equation}

Using eq. (\ref{eq:fsu2}) in eq. (\ref{eq:corte2}) and
remembering that $d = 2j+1$,
we have $\alpha_{j} = j$. As $2j+1$ is an integer,
this means that $j$ and consequently $\alpha_{j}$ can be an
integer or a semi-integer, as it is well-known. We can also
see that the lowest eigenvalue is $-j$ and the eigenstates 
can be written as $|j, j-m \rangle$, where $m$ goes from zero 
to $2j$. We should also say that the Casimir operator of $sl(2)$ 
is obtained from eq. (\ref{eq:casimir}). 

In general, in order to see the iterations through a graphical analysis
of the function $f$ we graph $y=f(x)$ together with
$y=x$. Where the lines intersect we have $x=y=f(x)$, so that
the intersections are precisely the fixed points. Now, for a
point $x_0$, different from the fixed point, in order to follow
its path through iterations with the function $f$, we perform
the following steps
\begin{enumerate}
\item
move vertically to the graph of $f(x)$,
\item
move horizontally to the graph of $y=x$, and
\item
repeat steps 1, 2, etc.
\end{enumerate}

We now present in fig. (1) a curious graphical representation of
a finite dimensional representation of the $sl(2)$ algebra where
the iterative aspect of this algebra is emphasized. In fig. (1) we
plot the function $f(\alpha) = \alpha - 1$ versus $\alpha$ for
$\alpha_{j} = j = 2$. We also plot the vertical line representing
the cut condition given by eq. (\ref{eq:corte2}), i.e.,
$\alpha_{j-d} = -\alpha_{j}-1=-j-1 = -3$.

\begin{figure}[h]
\centerline{\epsfig{file=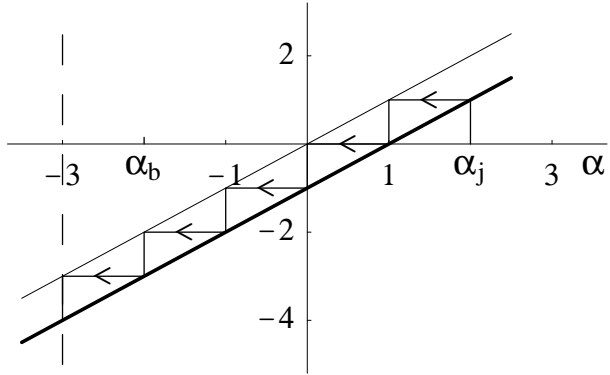,width=8cm}}
\caption{{\small Five dimensional representation of the $sl(2)$
algebra for $\alpha_{j} = 2$. The dashed line ($\alpha_{j-d} = -3$)
is the cut condition. The iterations are indicated by the arrows.
The thinner line is the function $f(\alpha) = \alpha$ and the thicker
one 
is the function $f(\alpha) = \alpha - 1$.} }
\label{fig1} 
\end{figure}

We see that the 
iterations of $j$ through $f$ reach exactly the intersection of
the vertical line given by the cut condition and the function itself.
If the starting point $\alpha_{j}$ is not an integer or semi-integer,
the future iterations of this value will never reach the
intersection with the cut line and the iterations will
evolve forever. There is no lower bound.
The dimension of this representation is infinite.
Note, also, that there is no fixed point in this case.
In this way, a 
graphical analysis of the function $f$ gives us quick and
useful information about the representation of the algebra,
without the need to realize an extensive calculation. 
It is also easy to see in
this graph that the eigenvalues always decrease in absolute value
when the iterations of the function $f$ increase.

\subsection{Linear deformations of $sl(2)$}\label{subsec.linear}

 Let us consider the function $f = r J_{0} - s$,
where $r$ and $s$ are real numbers.
 The eqs. (\ref{eq:alg1}-\ref{eq:alg3}) can be written as:
 \begin{eqnarray}
     \left[J_{0}, J_{-}\right]_{r} & = & -s J_{-}
     \label{eq:j-su2}  \\
     \left[ J_{0}, J_{+}\right]_{r^{-1}} & = & (s/r) J_{+}
     \label{eq:j+su2}  \\
     \left[J_{+}, J_{-}\right]\,\,\,\,\,\, & = &  (1-r^2) J_{0}^2 +
     (1 + 2rs -r) J_{0}
     + s (1-s) \, ,
     \label{eq:comutsu2}
 \end{eqnarray}
 where $[A,B]_{r} = AB - rBA$ is the $r$-deformed commutation of two
 operators $A$ and $B$.

 There are three cases of interest to be analysed: (I) $r=1$ and
 $s > 0; \, \, s \neq 1$, (II) $r>1$ and (III) $|r|<1$.
 In the first case we consider only
 positive (unlike $s=1$) values of $s$ because negative
 values of $s$ will not satisfy the condition given by
 eq. (\ref{eq:rangeb}). The $r$-commutator in
 eqs. (\ref{eq:j-su2}-\ref{eq:j+su2}) turns out to be
 standard commutator and the
eigenvalues of $J_{0}$ are
$\alpha_{j-m} = \alpha_{j} - ms$.
Once the dimension ($d$) of the representation is chosen,
the cut condition given by eq. (\ref{eq:corte2}) leads to the
following value of $\alpha_{j}$:
 \begin{equation}
     \alpha_{j} = \frac{sd-1}{2} \, .
     \label{eq:alfapossivel}
 \end{equation}
 For a fixed value of $s$ and for each different dimension of the
 representation we want, we have a different initial value allowed.
 As $s$ can be any positive real number ($\ne 1$), this implies
 that $\alpha_{j}$ can also be a real number.
 Note that the transformed operators $\tilde{J}_{\pm}=J_{\pm}/s$
and $\tilde{J}_{0}=J_{0}/s+(1-s)/(2s)$ obey the $sl(2)$ algebra,
 while $J_{\pm}$ and $J_0$ satisfy the
 algebra given by eqs. (\ref{eq:j-su2}-\ref{eq:comutsu2}) for
 $r=1$ and $s > 0; \, \, s \neq 1$.
 The graphical representation of this case is similar to the
 $sl(2)$ case, with constant spacing $s$ between the eigenvalues.
  
 In case (II), $r>1$ and thus exists a fixed point $\alpha^{*} = s/(r-1)$,
 where the function $f$ crosses the diagonal $y=\alpha$.
 This fixed point is unstable,
 ($(\partial f/\partial \alpha)|_{\alpha^{*}} = r >1$),
 showing that only points below $\alpha^{*}$ are allowed if we
 obey the condition given by eq. (\ref{eq:rangeb}). Also,
 eq. (\ref{eq:alfameio}) shows
 that $\alpha_{j} > -1/2$, a necessary condition to have
 $\alpha_{j} > \alpha_{j-d}$. Then, $s$ and $r$ should satisfy
 the inequality $\alpha^{*} > -1/2$ and $\alpha_{j}$ lies within
 $-1/2 < \alpha_{j} < \alpha^{*}$.  But even in this interval,
 only those values of $\alpha_{j}$ that satisfy eq. (\ref{eq:corte2})
 are allowed. 
 The eigenvalues of $J_{0}$ can be written as:
\begin{equation}
\alpha_{j-m} = f^m(\alpha_{j}) = r^m \alpha_{j} - s
\left[m\right]_{r} \, ,
\label{eq:alfamlinear}
\end{equation}
 where $\left[m\right]_{r} \equiv (r^m-1) / (r-1)$ is the Gauss number.
 The cut condition given by eq. (\ref{eq:corte2}) and the
 eq. (\ref{eq:alfamlinear}) yield us an expression for $\alpha_{j}$
 once the function $f$ has been given (i.e., that $r$ and $s$
 are given) and the dimension of the representation is chosen.
 We have for $\alpha_{j}$:
 \begin{equation}
     \alpha_{j} = (s [d]_{r} -1) / (r^d + 1) \, .
     \label{eq:alfajexpression}
 \end{equation}
 For each dimension we want, we have a different starting point,
 generally a real number.
 In fig. (2) we show an example of this
 case. 
 
 \begin{figure}[h]
\centerline{\epsfig{file = 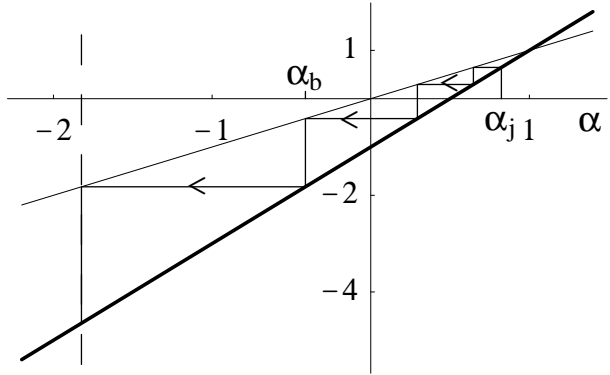,width=8cm}}
\caption{{\small Case II; 4-dimensional representation for $r=2$,
$s=1$, $\alpha_{j} = 14/17$ and $\alpha^{\star} = 1$. The dashed line
is the cut condition $\alpha_{j-d} = - 31/17$. The thinner line is
the function $f(\alpha) = \alpha$ and the thicker one
is the function $f(\alpha) = 2 \, \alpha - 1$.} }
\label{fig2} 
\end{figure}

There is also a marginal two dimensional representation for $r=-1$.
The case $r<-1$ is not allowed because it is not possible to obtain
a highest weight representation.
 
 In case (III), there is also a fixed point with the same
 formal expression for $\alpha^{*}$, but in this case $|r|<1$.
 This fixed point is stable
 ($\partial f/\partial \alpha)|_{\alpha^{*}} = r$; $|r| <1$,
 indicating that only the region with
 $\alpha_{j} > \alpha^{*}$ is
 allowed (since $\alpha_{j}$ is the highest value).
 The formal expressions given by eqs. (\ref{eq:alfamlinear}) and
 (\ref{eq:alfajexpression}) are still valid here, but with $|r|<1$.
 However, as in this case we
 are only considering $\alpha_{j} > \alpha^{*}$,
 the iterations of $f$ will approach the fixed point. Yet, note
 that in order to have a finite dimensional representation
 we must have $\alpha^{*} <  -\alpha_{j} - 1$, otherwise the dimension
 of the representation will be infinite.
 Also, under the restriction
 given by eq. (\ref{eq:alfameio}) yields, in this case,
 $\alpha^{*} < -1/2$.
 In this way, the 
 cut condition given by eq. (\ref{eq:corte2}) should be obeyed by
 the allowed values of $\alpha_{j}$. For a fixed function,
 there are infinitely
 countable possible values of $\alpha_{j}$, one for each
 respective dimension.
 A graphical representation of this case can be seen in
 fig. (3). 
 
 \begin{figure}[h]
\centerline{\epsfig{file = 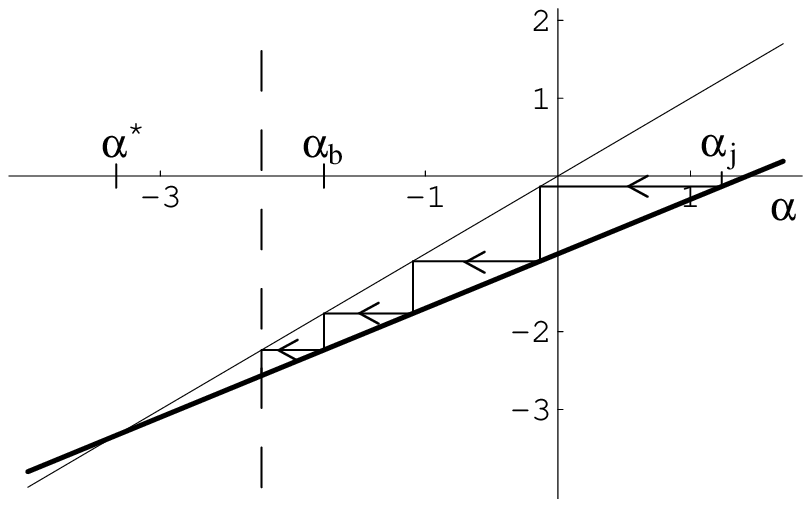,width=8cm}}
\caption{{\small Case III; 4-dimensional representation for $r=0.7$,
$s=1$, $\alpha_{j} = 1.23619\ldots$ and $\alpha^{\star} = -3.33333\ldots$.
The dashed line is the cut condition $\alpha_{j-d} = -2.23619\ldots$.
The thinner line is the function $f(\alpha) = \alpha$ and the thicker one
is the function $f(\alpha) = 0.7 \, \alpha - 1$.} }
\label{fig3} 
\end{figure}

 \subsection{Connection with $sl_{q}(2)$}\label{subsec.suq2}
 
 In this section we are going to show the connection of the
 generalized $sl(2)$ algebra for linear $f(J_{0})$,
 eqs. (\ref{eq:alg1}-\ref{eq:alg3}),
 with the deformation of $sl(2)$ found in the literature as
 $sl_{q}(2)$ \cite{algdeformed}.
 The $sl_{q}(2)$ generators, let us
 call them $S_{3}$ and $S_{\pm}$, have the following commutation
 relations among them \cite{algdeformed}:
 \begin{eqnarray}
    \left[S_{3}, S_{\pm}\right]  & = & \pm S_{\pm}
     \label{eq:s+}  \\
     \left[S_{+}, S_{-}\right]  & = & \left[2 S_{3}\right] \, ,
     \label{eq:s3}
 \end{eqnarray}
 where $[x] \equiv (q^{x} - q^{-x})/(q - q^{-1})$. The parameter
 $q$ is a real number and is called
 the deformation parameter of the algebra. When $q \rightarrow 1$
 the above commutation relations recover the $sl(2)$ relations.
 A simple transformation shows that $[x] = q^{-x+1} [x]_{q^2}$, where
 $[x]_{q^2} = (q^{2x}-1)/(q^2 - 1)$.
 The action of these generators to the states of an irreducible
 representation of the $sl_{q}(2)$ algebra, dimension of which
 is $2j+1$, 
 can be written as \cite{algdeformed}:
 \begin{eqnarray}
    S_{\pm} |j,j-m \rangle & = &
    \sqrt{q^{-2j+1}[j\mp (j-m)]_{q^2} [j \pm (j-m) + 1]_{q^2}} \,
    |j,j-m \pm 1 \rangle \nonumber \\
       S_{3} |j,j-m \rangle & = & (j-m) |j,j-m \rangle
     \label{eq:svec}
 \end{eqnarray}
 where $2j+1$ is a positive integer and $m = 0,\,1,\, 2,\, \ldots,\, 2j$.

In general, explicit functionals that map generators
of an specific algebra to another one can be found \cite{zachos}. 
In particular,
there is a specific map that convert the $sl(2)$ generators into
the generators of $sl_{q}(2)$ \cite{zachos}. The relevance of these
maps is that they may provide information on an unknown co-algebra
structure based on a known co-algebra.
 
In our case, if we remember the action of the $J_{0}$ generator in a
space 
of dimension $2j+1$, eq. (\ref{eq:j0m}), and use the expression of
$\alpha_{j-m}$ given by eq. (\ref{eq:alfamlinear}) with
$r \equiv q^2$, we see that expressing the generator $J_{0}$ as:

\begin{equation}
    J_{0} = q^{2(j-S_{3})} \alpha_{j} -
    s \left[j - S_{3}\right]_{q^2} \, ,
    \label{eq:j0s3}
\end{equation}
this generator acts to the $2j+1$ states of the representation of
the $sl_{q}(2)$ algebra exactly as it does on its own space of same
dimension. 

If we identify 
\begin{equation}
  J_{+} = \frac{
  \sqrt{(Q_{1}\alpha_{j} - Q_{2} [j-S_{3}+1]_{q^2})
  (Q_{3} \, \alpha_{j} + 1 + Q_{2} [j-S_{3}+1]_{q^2})}
               }
  {\sqrt{q^{-2j+1} [j-S_{3} +1]_{q^2} [j+S_{3}]_{q^2}}
  } \, 
  S_{+}
  \, ,
    \label{eq:j+s+}
\end{equation}
where $Q_{1} \equiv (q^2-2)/(q^2 -1)$, $Q_{2} \equiv (q^2 -1)\alpha_{j} -s$
and $Q_{3} \equiv q^2/(q^2 -1)$,
this generator also acts to the $2j+1$ states of the $sl_{q}(2)$
algebra exactly 
the same way it does to its own $2j+1$ space of states as given by
eqs. (\ref{eq:jm+})
and (\ref{eq:alfamlinear}). As $J_{-} = J_{+}^{\dagger}$,
the transformations given by eqs. (\ref{eq:j0s3}, \ref{eq:j+s+})
connect the $sl_{q}(2)$ algebra with the $r \neq 1$ linear case
of our formalism. 

Applying the same procedure just described above,
we can compute the inverse map, i.e., to express the $sl_{q}(2)$
generators in terms of $J_{(\pm,\, 0)}$, that could be used to obtain
some information on the co-algebra structure of the generalized $sl(2)$
algebra 
given by eqs. (\ref{eq:j-su2}-\ref{eq:comutsu2}).

Therefore, we have shown that this linear case is connected to
the $sl_{q}(2)$ algebra. Moreover, this formalism allows generalizations
of $sl(2)$ to more complex algebras obtained by considering
non-linear functions $f$ in eqs. (\ref{eq:alg1}-\ref{eq:alg3}). These
algebras, depending on the function $f$, will not simply be deformations
of $sl(2)$. 

\section{Non-linear functions}

In this section we consider some aspects of the
representation theory of
the algebra defined by eqs. (\ref{eq:alg1}-\ref{eq:alg3})
for $f(x) = t \, x^2 + r \, x - s$ .  In this case the algebra becomes
\begin{eqnarray}
\left[ J_0,J_+ \right]_{r^{-1}} &=& - r^{-1} \, (t J_{0}^2 - s) \, J_{+}
\;\;\; ,
\label{eq:comq1}\\
\left[ J_0,J_- \right]_{r} &=& J_{-} \, (t J_{0}^2 - s) \;\;\; ,
\label{eq:comq2}\\
\left[ J_+,J_- \right] &=& -t^2 J_{0}^4 -2trJ_{0}^3 +
(1-(1-s)t - r^2 + st)J_{0}^2 \\
\nonumber 
 & &  + (1 - r (1-2s))J_{0} + s(1-s)  \;\;\; .
\label{eq:comq3}
\end{eqnarray}
When $t=0$ we
recover the linear (or $r$-deformed) $sl(2)$ algebra given
in eqs. (\ref{eq:j-su2}-\ref{eq:comutsu2}).  For $t=0$ and
$r=s=1$ we recover the standard $sl(2)$ algebra.

We focus now on the analysis of
the finite dimensional representations of the
above quadratic $sl(2)$ algebra \footnote{From now one we are
considering $t>0$; the analysis of negative values of $t$ is similar.}.
To this aim we have to look for the finite dimensional representation
solutions to eqs. (\ref{eq:j0m}-\ref{eq:Tm}). Since we are starting
from 
a highest weight vector and in order to have a finite dimensional
representation we 
must find the conditions where the eq. (\ref{eq:jm-}) is
identically null. This is obtained by analysing the zeros of the equation
\begin{equation}
N_{d-1}^2 = (\alpha_{j} - \alpha_{j-d}) (\alpha_{j} + \alpha_{j-d} + 1)
= 0  \, .
\label{eq:nzero}
\end{equation}
In this case,
we can find solutions to eq. (\ref{eq:nzero}) in two different ways.
In the first manner,
as in the linear case, we find the solutions to $N_{d-1}=0$ that satisfy
the cut condition $\alpha_{j} + \alpha_{j-d} + 1=0$. In the non-linear
case
we can also find the zeroes of $N_{d-1}=0$ coming from $\alpha_{j} =
\alpha_{j-d}$.
The zeros of the term
$\alpha_{j} - \alpha_{j-d}$ can be obtained through the analysis and
the stability
of the fixed points of $f(x) = t \, x^2 + r \, x - s$
and their composed functions \cite{algebra2}.

In order to analyse the stability of the fixed points of $f(x)$
it is convenient to sort this analysis out in three cases: 
(I) $\Delta < 0$, (II) $\Delta = 0$ and (III) $\Delta > 0$, for
$\Delta = (r-1)^2 + 4\,t\,s$. In the first case there is no
fixed point and we see, by a graphical analysis similar to that
discussed in subsection \ref{subsec.su2}, that there is no
finite dimensional representation coming from $\alpha_j=
\alpha_{j-d}$; in case (II), we have one fixed point given by
$\alpha^{\star} = (1-r)/2t$. This fixed point corresponds to a
trivial one-dimensional representation of the algebra for
$\alpha_j = \alpha^{\star}$ since $N_0 = 0$
($\alpha_{j-1} = \alpha_{j} = \alpha^{*}$).

Case (III) is less trivial. In this case it is also possible to have
attractors of period 1, 2, 4, $\cdots$ and even a chaotic region
in the space of parameters ($t$, $r$, $s$, $\alpha_0$) where, as it is
well known, there are cycles associated with all integer numbers.

For $0 < \Delta < 4$ there is only trivial one-dimensional
representations associated to the fixed points
$\alpha^{\star} = f(\alpha^{\star})$, with highest weight:
\begin{equation}
 \alpha^{\star}_{\pm} = \frac{1-r \pm \sqrt{\Delta}}{2 \, t} \;\;\; .
\label{eq:alfastar}
\end{equation}
At $\Delta = 4$ the one-cycle looses stability and a stable two-cycle,
solution to $\beta^{\star} = f^2(\beta^{\star})$ ($f^2(x)\equiv
f(f(x))$), 
appears. The solutions to the two-cycle equation that are not
attractors of 
period 1 (fixed point of $f$) are
\begin{equation}
\beta^{\star}_{\pm} = \frac{-1-r \pm \sqrt{\Delta_1}}{2 \, t} \;\;\; ,
\label{eq:beta1}
\end{equation}
where $\Delta_1 = -3 - 2 \, r + r^2 + 4 \, t \, s$.
In this case, $\Delta > 4$, we have a two-dimensional representation
of the algebra simply by choosing the highest weight, $\alpha_j$,
as the highest element of the two cycle, i.e, $\beta^{\star}_+$.
Note that in eq. (\ref{eq:nzero}) the term $\alpha_{j} - \alpha_{j-d}$
becomes, in this two-dimensional case, $\beta^{\star}_+ -
f^2(\beta^{\star}_+)$, that is identically zero.
The matrix representation is given by
\begin{equation}
    J_{0}= \left(  
    \begin{array}{cc}
        \beta_+^{\star} & 0 \\
        0  & \beta_-^{\star}
    \end{array}
    \right)  , \hspace{0.1cm}
       J_{+}= \left(
    \begin{array}{cc}
        0 & 0  \\
        N_{0} & 0  
    \end{array}
    \right) , \hspace{0.1cm}
    J_{-} = J_{+}^{\dag} \hspace{0.1cm} ,
    \label{eq:matriz2}
\end{equation}
where $N_0$ is computed for $\Delta > 4$ and $\alpha_j = \beta^{\star}_+$.

For $\Delta > 6$ we will have other cycles, of
length 4, 8, \ldots, $2^k$ \ldots,  entering then the
chaotic region where all cycles will be present. In general,
for a $d$-cycle we have
a $d$-dimensional representation where the highest
weight of the representation is the largest element of
the cycle. The term $\alpha_{j} - \alpha_{j-d}$ of
eq. (\ref{eq:nzero}) is identically null for this cycle.

There are also finite dimensional representations
coming from the zeroes of the cut condition
$\alpha_{j} + \alpha_{j-d} + 1 = 0$ in the expression of $N_{d-1}$.
For example, the regions associated with the possible
highest weight vector solutions for the first cycles are better
understood by studying the corresponding $\Delta$ intervals.
For the one-cycle ($0 < \Delta < 4$), the following region  
\begin{equation}
\alpha_{-}^{*} < \alpha_{j} < \alpha_{+}^{*} \, ,
\label{eq:delta1}
\end{equation}
is the only region in the $\alpha$ real axis where it is possible to find
highest weight vectors since iterations of $\alpha$ in this region give
lower values
than the initial one.
In order to select $d$-dimensional representations we must pick up
the points in this interval that
satisfy the cut condition $\alpha_{j} + \alpha_{j-d} + 1 = 0$.
For example, $r=s=1$, $t=0.1$ ($\Delta =0.4$) with highest weight
$\alpha_{j}=0.476105$ is a possible solution to the cut condition
corresponding to a two-dimensional representation. Note also that
the highest weight of this two-dimensional representation is
within the $\alpha$-region for the one-cycle given in eq.
(\ref{eq:delta1}) since $\alpha_{-}^{*}=-3.16228=-\alpha_{+}^{*}$
for the values of the parameters under consideration.


For the two-cycle ($4 < \Delta < 6$), the following region
\begin{equation}
\beta^{\star}_+ < \alpha_j < \alpha^{\star}_+ \; ,
\label{eq:delta2}
\end{equation}
is the region in the $\alpha$ real axis where it is possible to find
highest weight vectors for $d$-dimensional representation for any
finite value of $d$, apart of the two-dimensional representation
($\alpha_{j} = \beta_{+}^{*}$).
In order to seek for finite dimensional representations we have to
find the points in this interval that satisfy
the cut condition. For example, in this region we have for the
parameters
$r=t=1$, $s=1.1$  ($\Delta=4.4$), giving
$\beta^{\star}_+ =-0.683772$ and $\alpha^{\star}_+ =1.04881$,
two possible two-dimensional representations with highest
weights given by $\alpha_j= \pm 0.316228$. 

For higher cycles the analysis is similar.
In all cycles the allowed regions to get possible
finite dimensional representations range from the largest
element of the cycle up to $\alpha_{+}^{*}$.



\newpage

\noindent
{\large \textbf{Acknowledgments}}

\vspace{0.5cm}

We would like to thank Pronex/MCT and CNPq for a partial 
support. M. A. Rego-Monteiro thanks the INFN section
of Torino, Italy, for a partial support and the dipartimento
di fisica dell' Universit\`{a} di Torino, Italy, where part
of this work was developed.

\end{document}